\documentclass[%
 reprint,            
 superscriptaddress, 
 amsmath,amssymb,    
 aps,                
 prd,                
 floatfix,           
]{revtex4-2}

\usepackage{graphicx}   
\usepackage{dcolumn}    
\usepackage{bm}         
\usepackage[colorlinks=true, allcolors=blue]{hyperref} 

\newcommand{\aapr}{Astron.\ Astrophys.\ Rev.}

\newcommand{\aj}{Astron.\ J.}
\newcommand{\apjl}{ApJ~Lett.}
\newcommand{\apjs}{ApJS}
\newcommand{\araa}{ARA\&A}
\newcommand{\apph}{Astropart.\ Phys.}
\newcommand{\jcap}{JCAP}
\newcommand{\mnras}{MNRAS}

\begin{document}

\title{Rare Transients in Nearby Galaxies Explain Ultra--high--energy Cosmic Rays}

\author{I. Bartos}
\email{imrebartos@ufl.edu}
\affiliation{Department of Physics, University of Florida, PO Box 118440, Gainesville, FL 32611-8440, USA}

\author{M. Kowalski}
\email{marek.kowalski@desy.de }
\affiliation{Deutsches Elektronen-Synchrotron DESY, Platanenallee 6, 15738 Zeuthen, Germany}
\affiliation{Institut für Physik, Humboldt-Universität zu Berlin, 12489 Berlin, Germany}

\begin{abstract}
The origin of ultra--high--energy cosmic rays remains one of the central open questions in astroparticle physics. 
Recent measurements reveal anisotropies in arrival directions, a rigidity--dependent composition dominated by intermediate--mass nuclei, and striking hemispheric differences in the energy spectra. Here we show that \emph{rare transients in nearby galaxies} can naturally account for these features. 
In our fiducial neutron--star merger model, the cosmic ray flux above $25$~EeV is dominated by ten nearby galaxies within $8\,$Mpc. This accounts for the observed hotspots: seven of the ten brightest galaxies coincide with reported excess regions, a chance probability of $p\sim0.001$. Nearby transients can also explain the spectral excess of TA over Auger and modify the rigidity--aligned succession of isotopes.
\end{abstract}

\maketitle


The nature and origin of cosmic rays remain among the longest-standing open questions in astrophysics \cite{2011ARA&A..49..119K,blasi2013}. Despite major experimental progress, the sources capable of accelerating particles to the highest observed energies have not been firmly identified. Charged cosmic rays are deflected by Galactic and extragalactic magnetic fields, scrambling their arrival directions and erasing temporal information, making direct source associations difficult. 

Ultrahigh-energy cosmic rays (UHECRs), typically defined as particles above $E \gtrsim 5 \times 10^{18}$\,eV (5\,EeV), offer a unique probe of this problem. Their extreme energies imply an extragalactic origin, since they cannot be confined by the Galactic magnetic field \cite{2011ARA&A..49..119K}. They serve as signposts of the most energetic cosmic processes, testing models of relativistic outflows and particle acceleration. 

During propagation, UHECRs lose energy through interactions with the cosmic microwave and infrared backgrounds, restricting their horizon to tens of megaparsecs at the highest energies \cite{greisen1966,zatsepin1966,2012APh....39...33A}. On such scales, the universe is highly inhomogeneous, raising the prospect of linking anisotropies in arrival directions to nearby structure \cite{2022MNRAS.510.1289V}. 

Observational progress in recent years has revealed a set of striking clues. The Pierre Auger Observatory (Auger; \cite{2015172}), which observes the Southern sky below a declination of $44.8^\circ$, has reported an excess region, or “hotspot”, above $38$\,EeV in the direction of Centaurus A \cite{2022ApJ...935..170A}. The Telescope Array (TA; \cite{ABUZAYYAD201287}), which observes the Northern sky above  $-16.0^\circ$, has identified two hotspots: one toward the Perseus--Pisces supercluster above $25$\,EeV \cite{TAhotspot25}, and another toward Ursa Major above $57$\,EeV \cite{ta2014hotspot57}.  In addition to these intermediate-scale excesses, Auger has measured a large-scale dipole above $\sim 8$\,EeV pointing away from the Galactic plane and roughly aligned with nearby large-scale structure, after accounting for deflections in the Galactic magnetic field\cite{Auger2017dipole,2024ApJ...975L..21B}.

Independent information comes from composition studies. Measurements of the depth of shower maximum, $X_{\mathrm{max}}$ show a trend toward heavier nuclei at the highest energies \cite{auger2017composition}. This behavior points to charge-dependent limits to particle acceleration. 
Persistent differences between the northern (TA) and southern (Auger) spectra add further complexity: the TA spectrum appears harder and cuts off at higher energy. Calibration alone cannot explain this difference \cite{Auger2020PRDSpectrum,TAspectrum2023}. Taken together, these observations suggest that the UHECR sky is structured, nearby, and of mixed composition \cite{Plotko:2022urd}.

Multiple source classes have been proposed to account for these signals, including active galactic nuclei (AGN), gamma-ray bursts, starburst galaxies, and galaxy clusters \cite{abraham2007agn,2015ApJ...804...15A,Waxman:1995vg,anchordoqui2019review}. 
While correlations between UHECRs and nearby AGN catalogs and starburst galaxies have been reported \cite{Aab_2018}, no single class has been firmly established. 

Theoretical work has long examined how magnetic propagation and local structure shape the observed flux \cite{Waxman:1996zn,2000JHEP...02..035H,2022MNRAS.510.1289V,2023MNRAS.524..631T}, and recent studies show that UHECR anisotropies can diagnose the role of rare transient sources and the intervening magnetic fields \cite{2024ApJ...975L..21B,2024ApJ...966...71B}.  
\citet{2024ApJ...972....4M} argued that if UHECRs originate from rare transients, only a small number of active hotspots should be visible at a given time. By modeling propagation through plausible magnetic environments, they showed that such transients remain observable for $10^4$–$10^5$\,yr, naturally producing long-lived anisotropies. 
Similarly, \citet{2024ApJ...962L...5U} traced the 244~EeV “Amaterasu” event to a nearby extragalactic transient, likely within the Local Sheet or Local Group, reinforcing the case for a transient origin of the highest-energy particles.


A common feature of many proposed \emph{steady} source classes, such as powerful AGN and massive galaxy clusters, is that they are sparse and typically distant, so that only a small number of systems lie within the UHECR propagation horizon. For starburst and star–forming galaxies the situation is more nuanced: while a few nearby examples (e.g.\ NGC~253, M82) lie within a few Mpc and can be individually important in anisotropy studies, the bulk of the population that carries most of the luminosity density is still at distances comparable to or larger than the attenuation length.

An alternative is that the accelerators are \emph{stellar transients} such as neutron star mergers \cite{2014PhRvD..89f3006T,2018ApJ...866...51K,2024arXiv240517409Z,2025PhRvL.134h1003F,2025arXiv250622625F}, tidal disruption events (TDEs; \cite{2014arXiv1411.0704F,PhysRevD.96.063007}), or collapsars \cite{Waxman:1995vg}.   
If such sources dominate, two consequences follow. First, these transients occur in nearly every galaxy, therefore, the local overdensity of galaxies implies that the UHECR sky is shaped by a few nearby systems. Second, their transient nature makes the flux from individual galaxies variable.

Motivated by these clues, we performed a population-level directional test of stellar transients as UHECR sources. To carry this out we assembled a galaxy catalog that enables a statistical search for correlations with neutron star mergers, TDEs, and core-collapse supernovae. To our knowledge no analogous catalog-based test has been carried out for stellar transients. 

The catalog assigns each galaxy a physically motivated weight based on its stellar mass ($M_\star$) and star-formation rate (SFR), which govern the expected occurrence of these transients. We based this catalog primarily on the \emph{HECATE} database \cite{Kovlakas2021HECATE}, which provides homogeneous $M_\star$ and SFR. To ensure full-sky coverage and accurate distances out to 100\,Mpc, we extended it with redshift-independent distances from \emph{Cosmicflows-4} \cite{CF4} and the \emph{Updated Nearby Galaxy Catalog} \cite{UNGC}, and with NIR/MIR photometry from \emph{AllWISE} \cite{allwise} and \emph{2MASS} \cite{Skrutskie2006TwoMASS,Jarrett2000XSC}. When HECATE values were unavailable, we derived $M_\star$ from WISE $W1$/$W2$ photometry \cite{Cluver2014WISEML} or from 2MASS $K_s$ magnitudes using the solar $K_s$ reference from \cite{Willmer2018SolarMags}, and estimated SFR from the empirical main-sequence relation between SFR and $M_\star$ at $z\simeq0$. For our fiducial model we adopted neutron star mergers, estimating their intrinsic rates from empirical scaling with $M_\star$ and SFR \cite{2020MNRAS.491.3419A} and normalizing to the local gravitational-wave-inferred rate \cite{2025arXiv250818083T} (Appendix~A).
\newline

\noindent{\bf Cosmic-ray emission and propagation.}--- 
We adopted the combined spectral and composition fit on cosmic rays above $10^{17.8}$\,eV detected by Auger~\cite{2017JCAP...04..038A}.  
Of the two local minima reported, we used the second, heavier--composition solution, which better reproduces the highest--energy data (our results are not sensitive to this choice). 
Propagation was simulated with \textsc{SimProp}~v2r4~\cite{Aloisio_2017}, which self-consistently computes energy losses from pair production, photopion production, photodisintegration, and cosmological redshift during UHECR propagation through the CMB and EBL. Sources beyond 100\,Mpc were approximated as an isotropic background (Appendix~B). 

Magnetic fields are uncertain and different plausible configurations yield similar broadening. For a convenient choice here we modeled cosmic–ray transport across three propagation regions mostly following \cite{2024ApJ...972....4M}: 
(i) an extragalactic magnetic field (EGMF), assumed homogeneous with strength $B_{\rm EGMF}=10^{-2}$\,nG and coherence length $\ell_{\rm c, EGMF}=1$\,Mpc; 
(ii) the turbulent magnetic field of the \emph{Local Sheet} (LS), represented as Kolmogorov turbulence filling a disk-like structure $10$\,Mpc in diameter and $0.5$\,Mpc thick, tilted by $8^\circ$ with respect to the supergalactic plane and nearly perpendicular to the Galactic plane \cite{2014MNRAS.440..405M}, with characteristic amplitude $B_{\rm LS}=10$\,nG and coherence scale $\ell_{\mathrm{c,LS}}=10$\,kpc; and 
(iii) the Galactic magnetic field (GMF), approximated by a turbulent component extending out to $10$\,kpc with amplitude $B_{\rm GMF}=1\,\mu$G and coherence length $\ell_{\rm c,GMF}=100$\,pc. 
Deflections were treated in the small--angle regime with rigidity scaling
$\theta \propto R^{-1}$, with rigidity $R \equiv E/Z$ for a
particle of energy $E$ and charge $Z$. For each source we use this scaling to compute the broadening from the extragalactic
magnetic field across the relevant energy band and mass composition, and we denote
the resulting band--averaged contribution as $\theta_{\rm EGMF}(E)$ (its underlying
dependence remains $\propto Z/E$). The contributions from the Local Sheet and
Galactic fields are computed with the same rigidity scaling, but since they act
only along short path lengths near the observer and vary weakly across our narrow
energy bands, we evaluate them at a representative band--averaged rigidity
$\bar R$ and treat them as constants within each band. The net rms deflection is
then $\theta_{\rm rms}(E)
=
\bigl[
  \theta_{\rm EGMF}^2(E)
 +\theta_{\rm LS}^2(\bar{R})
 +\theta_{\rm GMF}^2(\bar{R})
\bigr]^{1/2}$. 
Each source’s flux was then convolved with a von~Mises–Fisher kernel, with dispersion $\theta_{\rm rms}(E)$, representing the angular probability distribution of arrival directions.
\newline

\noindent{\bf Nearby galaxies dominate the UHECR flux.}---
We found that at energies $\gtrsim 32$\,EeV (the threshold used in Auger's recent anisotropy searches; \citealt{2022ApJ…935..170A}) the top ten brightest galaxies contribute a large fraction of the cosmic ray flux in the neutron star merger scenario. Here we treated the M81 group as a single system due to its compact membership. Table \ref{tab:top10} shows the top ten galaxies and their contribution for $>32\,$EeV (see Table \ref{tab:top10_new_60} for $>60\,$EeV).

In addition to neutron star mergers, we considered luminosities proportional to SFR (as expected for CCSNe or collapsars), to stellar mass (as a proxy for populations of compact remnants), and to a simplified TDE rate model (Appendix A.4). For the neutron star merger, SFR and stellar mass cases the ten brightest nearby galaxies remained largely the same. For TDEs, half of the galaxies differed, and in particular the M81 group contributed little, making TDEs less suitable to explain the $57\,$EeV TA hotspot (Appendix~C).

The dominance of nearby galaxies arises from two effects acting in concert. First, the short attenuation lengths of UHECRs at the highest energies means that nearby sources have outsize contributions. Second, the source population contains a pronounced overdensity of matter in the vicinity of the Milky Way (see Fig.~\ref{fig:cdf_merger}). 
\newline

\begin{table}[t]
\centering
\caption{Fractional contributions (\%) of the ten brightest galaxies in our fiducial neutron star merger model to the
all--sky UHECR flux ($>32$\,EeV) in the allsky, Auger--weighted, and TA--weighted skymaps. 
Also shown are associated hotspot (if any) with energy thresholds.}
\label{tab:top10}
\begin{tabular}{l|cccccc}
\hline\hline
Galaxy       & Dist. & Allsky & Auger & TA & Hotspot & $E$ \\
             & [Mpc] & [\%]   & [\%]  & [\%] &        & [EeV] \\
\hline
Andromeda    & 0.8  & 20   & 0    & 34   & TA-PPSC    & $>25$ \\
M81 group    & 3.6  &  7   & 0    & 14   & TA-Ursa    & $>57$ \\
NGC 253      & 3.5  &  3   &  9   &  0   &            &       \\
NGC 5128     & 3.5  &  2   &  6   &  0   & Auger      & $>38$ \\
NGC 4945     & 3.6  &  2   &  6   &  0   & Auger      & $>38$ \\
M33          & 0.9  &  2   &  0   &  2   & TA-PPSC    & $>25$ \\
IC342        & 3.4  &  1   &  0   &  2   &            &       \\
ESO097-013   & 4.2  &  1   &  3   &  0   & Auger      & $>38$ \\
M87          & 16.4 &  1   &  1   &  1   &            &       \\
M32          & 0.8  &  1   &  0   &  1   & TA-PPSC    & $>25$ \\
\hline
Combined     &      & 38   & 25   & 56   &            &       \\
$d>100$\,Mpc &      & 27   & 23   & 11   &            &       \\
\hline\hline
\end{tabular}
\end{table}

\noindent{\bf Hotspots coincide with the brightest galaxies.}---
Out of the ten dominant galaxies, seven fall within the hotspot masks of TA and Auger. Andromeda, M32 and M33 lie in TA’s Perseus--Pisces hotspot above 25~EeV ($20^\circ$ radius) \cite{TAhotspot25,PhysRevD.108.103008}; the M81 group coincides with TA’s Ursa Major hotspot above 57~EeV, taken as the union of two published locations ($20^\circ$ radius each) \cite{ta2014hotspot57,2016PhRvD..93d3011H}; and Cen~A, NGC~4945, and ESO097-013 fall within Auger’s Centaurus hotspot above 38~EeV ($27^\circ$ radius) \cite{2022ApJ...935..170A}. 

Comparing this overlap to a null hypothesis in which hotspot centers are re-sampled according to the experiment's directional exposure (uniform in right ascension and with declination dependence computed following Sommers \cite{Sommers2001}), while keeping galaxy positions fixed, yields a chance probability of $p\simeq0.003$. If weighing by predicted flux, we obtain $p\simeq0.002$. If we use SFR and stellar-mass weights we obtain similar p-values, while for TDEs, the p-values are larger (0.04/0.008).

Auger further reports an excess compatible with NGC253 above 40\,EeV within a $25^\circ$ window \cite{2022ApJ...935..170A}. If we treat this as a hotspot, the chance probability for neutron star mergers becomes $0.001$.

While the positional overlap is strong, our fiducial configuration, without accounting for temporal variation, overpredicts the UHECR fluxes inside the hotspots, indicating that joint modeling of flux, spectrum, and direction will further refine the picture. We also note that our $p$--value uses only the positional overlap between galaxies and the published hotspot masks. Coherent GMF bending is not modeled and is treated as 
an isotropic broadening; expected deflections at these energies are below the 
$20^\circ$--$27^\circ$ hotspot radii, so they are unlikely to change inclusion 
in a mask, though a full flux prediction would require a more detailed GMF 
treatment.
\newline

\noindent{\bf Magnetic delays set transient visibility.}---
For nearby sources, the observed duration of UHECR emission is set almost entirely by propagation through the Local Sheet magnetic field, with negligible contribution from the weak extragalactic void field or the thin Galactic layer. In the small-angle scattering regime, the rms delay can be estimated as $\Delta t_{\rm rms}\sim d^2 Z^2 B_{\rm LS}^2 \ell_{\mathrm{c,LS}}ce^2 /(4E^2)$,
for $c$ is the speed of light and $e$ is the electron charge \cite{Waxman:1996zn}. In the case of Andromeda at $d=0.8$\,Mpc, $E=25$\,EeV and $Z=7$ give $\Delta t\simeq  10^{4}$\,yr. 
An equivalent estimate follows from the angular deflection resulting in a hot-spot, 
$\Delta t_{\rm hs}\simeq d\theta_{\rm rms}^2/4c$, with a hotspot size of a few degrees yielding a consistent result.

The inferred signal duration can be compared with expected average time between two events in 
Andromeda+M32+M33 combined: $\sim3\times10^{4}$\,yr for neutron star mergers, 
$\sim3\times10^{3}$\,yr for TDEs, 
and $\sim10^{2}$\,yr for CCSNe. With a duration of $\sim10^{4}$\,yr, 
frequent classes (CCSN, TDE) would yield effectively continuous emission, whereas rarer channels 
(neutron star mergers, collapsars) produce intermittent fluxes.

\begin{figure}[t]
  \centering
  \includegraphics[width=1\linewidth]{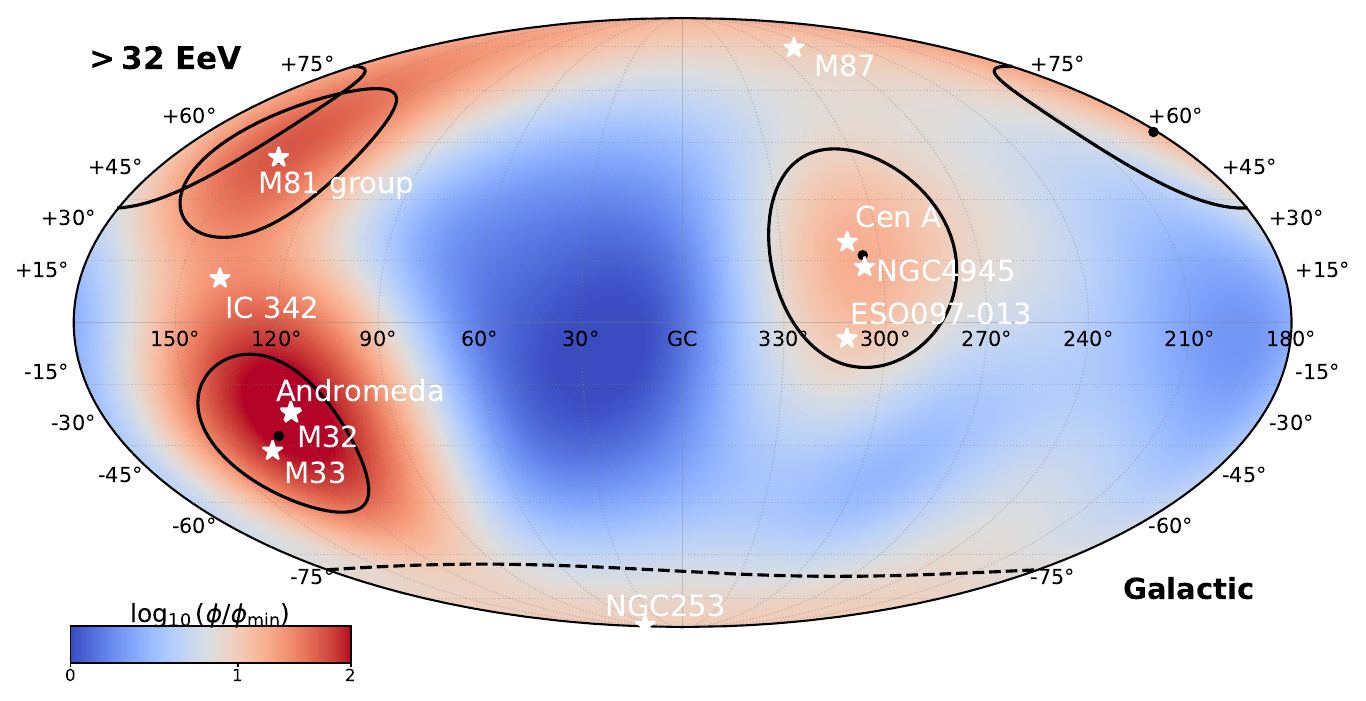}  \\
  \vspace{-0.5cm}
\caption{
Predicted UHECR skymap for $E>32$~EeV for our fiducial neutron star merger model. The locations of the top ten brightest galaxies are marked. Also shown are TA's Perseus--Pisces hotspot ($E>25$~EeV; near Andromeda; $20^\circ$ radius \cite{TAhotspot25}), Auger's Cen A hotspot ($E>38$~EeV; near Cen A; $27^\circ$ radius; \cite{2022ApJ...935..170A}), TA's Ursa Major hotspot ($E>57$~EeV; near M81; $20^\circ$ radius; both from \cite{ta2014hotspot57} and \cite{2016PhRvD..93d3011H}), and Auger's excess around NGC253 (dashed circle; $E>40$~EeV; $25^\circ$ radius \cite{2022ApJ...935..170A}).
}
\label{fig:transient_hotspots}
\end{figure}


This framework can explain why the Andromeda/M32/M33 direction can dominate the TA hotspot at $>25$\,EeV (Fig.~\ref{fig:transient_hotspots}), while fading at $>57$\,EeV, allowing the more persistent M81 group (with $\times 20$ longer duration) to emerge as the $>57$\,EeV hotspot.
\newline

\begin{figure}[t]
  \centering  \includegraphics[width=0.98\linewidth]{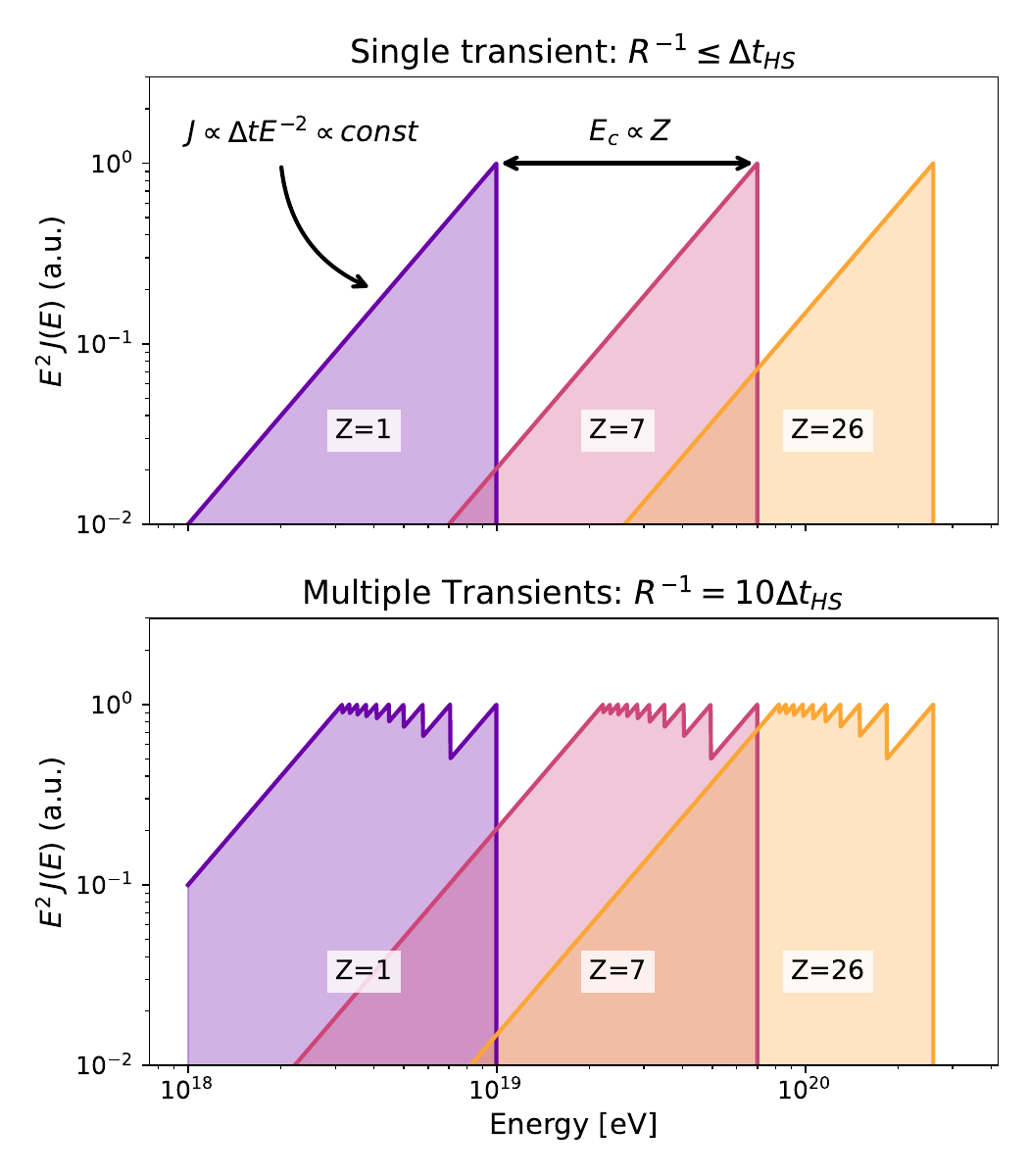}
  \vspace{-0.4cm}
  \caption{Illustration of the transient transfer effect. 
The top  panel illustrates the effect of a nearby transient source, with an intrinsic source spectrum following   $E^{-2}$ and a uniform composition $Z=(1,7,26)$. The observable arrival time falls within a duration window $\Delta t \propto (Z/E)^2$, and the flux is suppressed by the inverse of this duration. Once the arrival time is outside this window, the flux drops to zero. The maximum observable energy is rigidity-dependent (see Eq.\ \ref{eq:E_c}). The bottom panel shows the case where multiple transients contribute to a hotspot. 
While the peaks are smeared out, a rigidity dependent features can persist for rates that are not too large.} \label{fig:transfer_toy}
\end{figure}

\noindent{\bf Rigidity cutoff can arise from transient timing.}---
The effect of magnetic fields of cosmic rays with energy $E$ and nuclear charge $Z$ can be described as a magnetic transfer function that operates in rigidity space: particles with the same $E/Z$ experience the same magnification or suppression. Consequently, features observed for protons at energy $E$ reappear for nuclei of charge $Z$ at energy $ZE$, leading to the observed succession of isotopes aligned in rigidity.  Fig.~\ref{fig:transfer_toy} illustrates how a simplified transient transfer function modifies an injected spectrum.

The first consequence of this transfer function is an effective rigidity cutoff. Because the spread of magnetic delays scales as $\Delta t\propto (Z/E)^2$, protons arrive in short, sharp bursts \cite{Waxman:1996zn}, while heavier nuclei are dispersed over longer durations. Therefore, for sufficiently high energies, UHECRs from a
transient event may arrive and fade before the present epoch. In this case the observer misses the high-energy tail, producing a sharp cutoff in rigidity unrelated 
to the source’s intrinsic acceleration limit. 

The rigidity cutoff can be quantified by combining the rms deflection angle,
$\theta_{\rm rms} \propto Z B_{\rm LS} \ell_{\rm c,LS}^{1/2} d^{1/2} E^{-1}$,
with the small-angle delay relation,
$\Delta t \simeq d\,\theta_{\rm rms}^2/(4c)$,
\begin{equation}
\frac{E_c}{Z} \simeq 2\,\mathrm{EeV}
\left(\frac{d}{0.8\mathrm{Mpc}}\right)
\left(\frac{B_{\rm LS}}{\mathrm{10nG}}\right)
\left(\frac{l_{\rm c,LS}}{\rm 10 kpc}\frac{\mathrm{3\times10^4 yr}}{\Delta t}\right)^{1/2}
\label{eq:E_c}
\end{equation}
Here, $\Delta t$ is the time since the transient, which we approximated by the mean waiting time between transients. The relation illustrates that the effective maximum rigidity can be set by the interplay
of magnetic delays and source rates, not only acceleration physics. 
We note that the accelerator’s intrinsic maximum rigidity already imposes a baseline cutoff, but magnetic delays from nearby transients can further modify its high-energy 
end in a sightline--dependent way.
\newline

\noindent{\bf Rigidity--aligned composition.}---
A baseline rigidity ordering of the composition is expected from intrinsic 
acceleration physics. The rigidity-dependent magnetic delays discussed above 
add a further effect at the highest energies: because the delay scales as 
$(Z/E)^2$, nearby transients selectively shift the energies at which different 
species arrive, reshaping the observable succession of isotopes along a given 
sightline. Among the species that remain above the effective cutoff, the steeply 
falling spectrum ensures that the lightest surviving nucleus dominates the flux. 
Therefore, the most probable observational outcome for nearby transients is an 
enhancement of intermediate or heavy nuclei relative to protons. This trend is 
broadly consistent with observations: Auger measurements show that the 
composition becomes progressively heavier above $10$\,EeV, with strongly 
suppressed proton fraction \cite{auger2017composition}.

This rigidity-dependent composition pattern does not require a single dominant transient.
If several bright sources contribute within overlapping magnetic-delay windows, each will imprint a similar rigidity ordering but with slightly offset charge–energy boundaries.
The combined signal will still show a narrow sequence of dominant species, producing an overall composition that remains aligned in rigidity, even as the underlying transients vary.
This behavior is illustrated in the bottom panel of Fig.~\ref{fig:transfer_toy}, where ten simultaneous transients yield a broadened yet coherent rigidity-dependent composition feature.

One additional consequence of the magnetic transfer function is spectral hardening: if the observable arrival time falls within a duration window $\Delta t \propto (Z/E)^2$, the flux is suppressed by $\Delta t^{-1}$. Since this suppression decreases with energy, the observed cosmic ray spectrum hardens compared to the source spectrum.
\newline

\begin{figure}[t]
  \centering
  \includegraphics[width=1\linewidth]{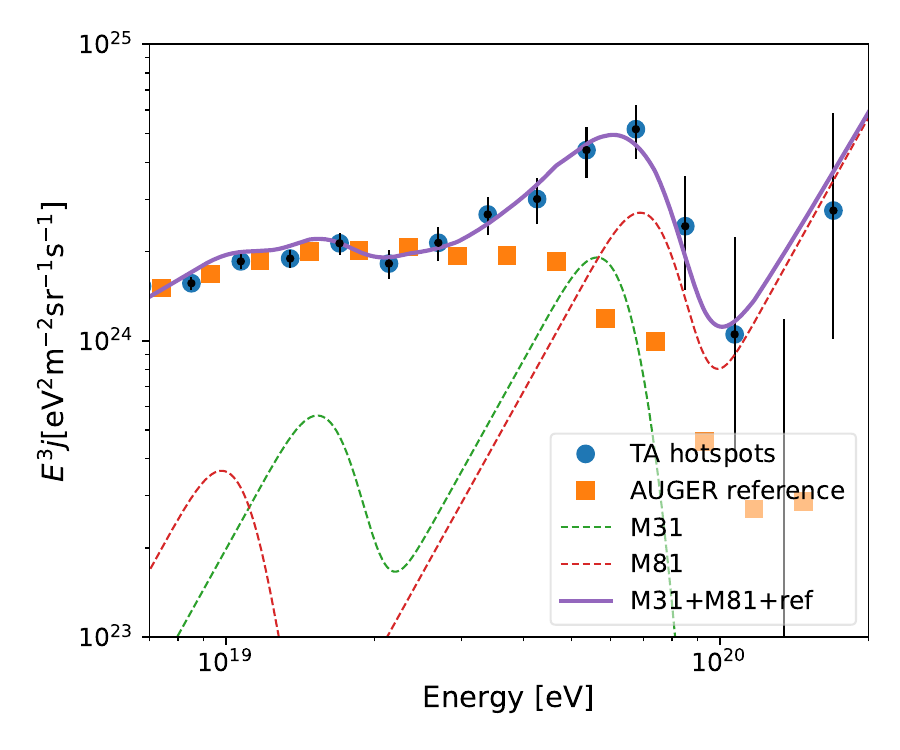}  \\
\vspace{-0.5cm}
\caption{
Spectrum from the combination of the two TA hotspots (inside the respective aperture of 25 and 20 deg), as shown in \cite{Kim:2025qmo}. The southern sky average from Auger is also shown. Also shown is our flux model for Andromeda (M31) and M81, as well as the sum of the two with 0.75 of the Auger reference flux added in addition (see text for details). }
\label{fig:hotspot_spectra}
\end{figure}

\noindent{\bf Northern spectra boosted by local anisotropy.}--- Fig.\ \ref{fig:hotspot_spectra} shows the combined observed spectrum of the two TA hotspots \cite{Kim:2025qmo}, along with Auger's sky-averaged spectrum as reference (the energy scales of Auger and TA were each rescaled by $\pm 4.5\%$, e.g.\ \cite{2025arXiv250905530B}). We modeled the expectations from Andromeda and the M81 group assuming an injection spectrum $J\propto E^{-2}$, and a uniform composition consisting of $Z=\left(1,7,26\right)$. We have kept the fiducal values for the magnetic field and coherence length in Eq.\ \ref{eq:E_c} but adjusted $\Delta t=1/6\times 10^5 \rm{ yr}$ (still within a factor of 2 of $\Delta t$ for Andromeda for the neutron star merger model) to match the observed cut-off energies (noting that there is a degeneracy between these parameters). The resulting maximal rigidity for Andromeda and M81 is then 2.5 and 10 EV, respectively. In our model, the flux dependence (up to the cut-offs) scale as $(E/Z)^2$. Furthermore, the resulting spectra were folded with a Gaussian of width 0.2 in $\log_{10}E$. The  difference in the spectral peaks observed for Andromeda and M81 is explained by the distance dependent maximal rigidity (Eq.~\ref{eq:E_c}), while the flux normalizations were chosen to match the data.

The model parameters are neither unique nor significantly fine-tuned, e.g., a similar match to the data could be obtained with a lighter composition consisting of $Z=\left( 1,2,7\right)$. And yet they capture key characteristics of the data, namely a hard spectrum and a sharp cut-off.

At $\sim 60$EeV, the combined flux inside the two hotspots is 3.5 times larger than the reference flux, or more than 50\% of the total flux from the  Southern hemisphere at these energies. This roughly matches the expectations obtained from our simulations (e.g.,
see Tab.\ \ref{tab:top10}). 

Next we look at the implied energetics of such a large flux. If one attributes the flux of the hotspots to two individual transients, the observed luminosity radiated over $\Delta t=1/6 \times 10^5$yr corresponds to $E_{\rm M31}\sim 3 \times 10^{50}{\rm erg}$ and $E_{\rm M81}\sim 5 \times 10^{51}{\rm erg}$, both per species and decade in energy.  The energetics derived for M31 appear consistent with models for neutron star mergers \cite{2025PhRvL.134h1003F} and long GRBs \cite{Waxman:1995vg}. For the M81 group, where a larger energy is required to compensate for the larger distance, the  energetics exceed expectations. An explanation could be that the transient has occurred more recently,  either by chance or because the transient rate in the M81 group exceeds that of Andromeda. This would not be surprising as the expected neutron star merger rate in the M81 group is about $\times 7$ higher than Andromeda. M81's higher implied rigidity is also in line with a more recent transient.    

The discussion of the Andromeda/M81 hotspots above illustrates that  rigidity dependent signatures are expected for distances up to a few Mpc, and thus how transients within the nearby accumulation of galaxies could shape the observable cosmic ray spectra.

A similar rigidity-ordered succession of elements arises in the classical \emph{Peters cycle} \cite{1961NCim...22..800P}, where all nuclei share a common maximum rigidity in their acceleration or confinement process. In that scenario, the composition becomes heavier with energy because lighter species reach their acceleration limit first. Our mechanism complements this picture: magnetic delays from nearby transients 
can shift and reshape the high-energy end of this intrinsic rigidity sequence in 
a direction-dependent manner.

A key observational distinction between the Peters-cycle scenario and the 
transient-delay picture lies in the source-to-source variability. In the Peters cycle, identical accelerators with a common maximum rigidity produce uniform spectra, showing the same rigidity-ordered cutoffs in all directions.
In contrast, for transient sources the effective cutoff depends on the time delay since the event and on the magnetic path length, so different sightlines can exhibit different cutoff rigidities.
Furthermore, because younger transients are observed earlier in their magnetic-delay sequence, their apparent luminosities are higher and their spectra extend to larger rigidities, leading to a natural correlation between luminosity and maximum observed energy that is absent in the Peters-cycle picture.
\newline




\noindent{\bf Conclusion and Outlook.}---
In summary, we find that rare, stellar transients in nearby galaxies can explain the hemispheric flux difference, hotspot locations, energy spectra, and composition trends. 
A handful of galaxies within $\lesssim 5$\,Mpc dominate the flux above tens of EeV, with a statistically significant overlap with TA and Auger hotspots; the spectral excess of TA over Auger, and its extension to higher energies, are likewise reproduced by the dominant role of Andromeda in the northern sky. 
The 25/57~EeV difference in TA is naturally accounted for by transient variability in Andromeda. 
Because magnetic lensing acts in rigidity space, it yields both the narrow dominance of individual species and the rigidity-aligned succession of isotopes. 

This framework yields concrete, testable predictions: (1) the UHECR flux from Andromeda and other nearby sources should be skewed toward heavier nuclei compared to more distant galaxies, reflecting its transient phase. (2) Auger should eventually confirm a second hotspot near NGC~253, consistent with the predicted flux hierarchy of the nearest galaxies. (3) The maximum observed rigidity should differ across directions, increasing for sightlines dominated by younger transients observed earlier in their magnetic-delay phase. Such a directional rigidity–luminosity correlation is not expected for the Peters-cycle. (4) With more data and improved composition information, proton hotspots should  appear at lower energies, that is, on the lowest rung of the rigidity ladder, in the same locations where hotspots are currently observed. (5) The observed cosmic ray composition and spectrum provide an inadequate representation for the larger Universe. Predictions based on extrapolations, such as those made for the cosmogenic flux of UHE neutrinos or photons, should be revisited.  

These signatures are within reach of forthcoming AugerPrime and TA$\times$4 measurements, offering decisive tests of the nearby-transient origin of UHECRs.
\newline 



\noindent{\bf Acknowledgements.}---
The authors would like to thank Teresa Bister, Francis Halzen, Kohta Murase, Anna Nelles, Andrew Taylor and Michael Unger for their valuable comments and suggestions, and the Telescope Array Collaboration for providing their observed spectrum within their two hotspot. I.B. acknowledges support from the National Science Foundation under Grant No. PHY-2309024. The authors used OpenAI’s ChatGPT \citep{openai2022chatgpt} during the preparation of this manuscript.

\appendix
\setcounter{table}{0}
\renewcommand{\thetable}{A\arabic{table}}
\setcounter{figure}{0}
\renewcommand{\thefigure}{A\arabic{figure}}
\makeatletter
\def\figurename{Fig.} 
\makeatother

\section{Galaxy catalog assembly}

\subsection{Stellar mass and star-formation rates from HECATE and ancillary catalogs}

We constructed a distance-limited galaxy catalog out to $d_{\max}=100$~Mpc, centered on the \emph{HECATE} v1.1 database \cite{Kovlakas2021HECATE}, which provides homogenized measurements of stellar mass, star-formation rate (SFR), and distance for nearby galaxies. When available, we adopted HECATE’s stellar mass, SFR, and distance ($D$, with quoted uncertainties $D_{\rm LO68}$/$D_{\rm HI68}$, etc.) directly.

To achieve full-sky coverage and supplement galaxies missing from HECATE, we extended the compilation using positions, distances, and photometry from \emph{Cosmicflows-4} (CF4) \cite{CF4}, the \emph{Updated Nearby Galaxy Catalog} (UNGC) \cite{UNGC}, and NIR/MIR photometry from \emph{AllWISE} \cite{allwise} and \emph{2MASS XSC} \cite{Skrutskie2006TwoMASS,Jarrett2000XSC}. When a HECATE distance was unavailable, we used CF4 redshift-independent distances or distance moduli with quoted uncertainties, and within the Local Volume we preferred UNGC distances based on tip-of-the-red-giant-branch (TRGB) or surface-brightness-fluctuation (SBF) indicators.

AllWISE provided profile-fit $W1$/$W2$ (Vega) magnitudes, while 2MASS XSC supplied $K_s$ (Vega) magnitudes as a fallback. We cross-matched galaxies to AllWISE using a $6''$ radius and to 2MASS XSC using $30''$. When both CF4 and AllWISE photometry were available, we retained CF4 values unless missing.

When HECATE stellar masses were unavailable, we derived $M_\star$ from WISE photometry following the color-dependent mass-to-light calibration of \citet{Cluver2014WISEML},
\begin{align}
\log_{10}\!\left(\frac{M_\star}{L_{W1}}\right) &= B_0 + B_1\,(W1 - W2),\\
\log_{10}\!\left(\frac{L_{W1}}{L_{\odot,W1}}\right) &= -0.4\,(M_{W1} - M_{\odot,W1}),\\
M_\star &= \left(\frac{M_\star}{L_{W1}}\right)\,L_{W1}.
\end{align}
where $M_{W1} = W1_{\mathrm{Vega}} - \mathrm{DM}$, $M_{\odot,W1} = 3.24$~mag (Vega), and $(B_0, B_1) = (-0.376, -1.053)$.
If only $W1$ was available, we estimated $M_\star$ from the luminosity relation \cite{Jarrett2023}:
\begin{align}
\log_{10} M_\star &= A_0 + A_1\,\log_{10} L_{W1} \nonumber\\
&\quad + A_2\,\big(\log_{10} L_{W1}\big)^2 
      + A_3\,\big(\log_{10} L_{W1}\big)^3 .
\end{align}
with $(A_0, A_1, A_2, A_3) = (-12.6, 5.0, -0.44, 0.02)$.
If WISE was unavailable but 2MASS $K_s$ existed, we adopted $M/L_{K_s}=0.6$ and $M_{\odot,K_s}=3.28$~mag \citep{Willmer2018SolarMags}.

For galaxies missing SFR estimates in HECATE, we inferred SFRs from the local star-forming main sequence, $\log_{10},\mathrm{SFR} = 0.76,\log_{10}M_\star - 7.64$, consistent with nearby-galaxy observations.

\subsection{Neutron star merger rate}

For each galaxy we estimated the intrinsic neutron star merger rate using the
local ($z\!\approx\!0$) scaling proposed by \citet{2020MNRAS.491.3419A},
which relates the specific rate to stellar mass and star-formation rate (SFR):
\begin{equation}
\label{eq:artale}
\begin{aligned}
\log_{10}\!\left(\frac{n_{\rm GW}}{\mathrm{Gyr}^{-1}}\right)
&= \beta_1\,\log_{10}\!\left(\frac{M_\star}{M_\odot}\right)
\\[2pt]
&\quad + \beta_2\,\log_{10}\!\left(\frac{\mathrm{SFR}}{M_\odot\,{\rm yr}^{-1}}\right)
+ \beta_3 .
\end{aligned}
\end{equation}
with $(\beta_1,\beta_2,\beta_3)=(0.800,\,0.323,\,-3.555)$.

To make this rate consistent with the most recent gravitational-wave constraints, we applied a uniform normalization factor so that the
catalog-implied density matches a chosen target local neutron star merger rate density,
$\mathcal{R}_{\rm target}$. For the target rate we adopted the mean merger rate value of $49$\,Gpc$^{-3}$yr$^{-1}$ estimated using the weakly model-dependent Binned Gaussian Process model from the most recent gravitational wave catalog GWTC-4 \cite{2025arXiv250818083T}.

\subsection{Cumulative merger rate with distance}

\begin{figure}[t]
  \centering
  \includegraphics[width=0.73\linewidth]{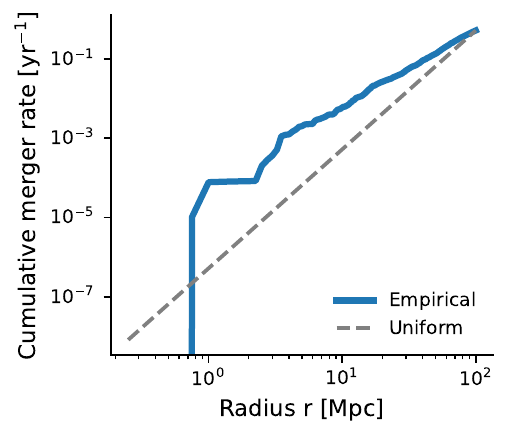}
  \vspace{-0.3cm}
  \caption{Cumulative distribution of the predicted neutron star merger rate with distance from Earth (solid line), compared to the expectation from a homogeneous source distribution (dashed line). The local overdensity produces a steep rise within the nearest tens of Mpc, implying that nearby galaxies dominate the expected UHECR flux.}
  \label{fig:cdf_merger}
\end{figure}

To illustrate the role of the local overdensity, in Fig.~\ref{fig:cdf_merger} we show the cumulative distribution of the predicted neutron star merger rate as a function of distance from Earth. For comparison, we also show the expectation for a homogeneous source distribution. The catalog indicates a much steeper rise within the local $\sim 5$~Mpc than the homogeneous case, demonstrating that the local overdensity strongly enhances the contribution of nearby galaxies to the observed UHECR flux.

\subsection{Tidal disruption event rates}
\label{app:TDE}

For each galaxy in our catalog we estimated a baseline TDE rate using simple scaling relations with black hole mass. In the absence of high-resolution nuclear stellar profiles for most galaxies, we adopted a commonly used prescription 
\cite{2016MNRAS.455..859S} in which the per-galaxy rate decreases weakly with black hole mass, with a sharp cutoff at high masses where main-sequence stars are swallowed whole. Specifically, we assumed
\begin{equation}
    \dot{N}_{\rm TDE} \;=\; 2\times10^{-4}\,{\rm yr}^{-1}\,
    \left( \frac{M_{\bullet}}{10^{6}\,M_\odot} \right)^{-0.25},
\end{equation}
for $M_{\bullet} < 3\times10^{8}\,M_\odot$, and $\dot{N}_{\rm TDE}=0$ otherwise. 

Black hole masses were estimated from the stellar mass of each galaxy via a scaling relation $M_{\bullet} \simeq 10^{-3}\,M_\star$ \cite{KormendyHo2013}, which provides a rough mapping appropriate for population-level studies. This yields characteristic rates of $\sim10^{-4}\,\mathrm{yr}^{-1}$ for galaxies hosting $10^{6}\,M_\odot$ black holes, declining slowly toward higher masses and vanishing above the swallowing threshold.

To avoid assigning unrealistically high weights to very low-mass galaxies, we further applied a black hole \emph{occupation fraction} $f_{\rm occ}(M_\star)$ that down-weights or removes systems unlikely to host a central massive black hole. The occupation fraction was modeled as a smooth, monotonic function of stellar mass,
increasing from $f_{\rm occ}\!\simeq\!0.02$ at $M_\star\!=\!10^8\,M_\odot$ to unity above $10^{10}\,M_\odot$, consistent with empirical estimates for local dwarf and spiral galaxies \cite[e.g.,][]{ReinesVolonteri2015}. We also set $f_{\rm occ}=0$ for galaxies whose inferred black hole masses fall below $M_{\bullet}<10^{5}\,M_\odot$, where central black holes are not expected to form or persist. The effective per-galaxy rate was thus
\begin{equation}
    \dot{N}_{\rm TDE,eff} \;=\; f_{\rm occ}(M_\star)\,\dot{N}_{\rm TDE}(M_\bullet).
\end{equation}
This treatment suppresses contributions from small galaxies while preserving the expected behavior at higher masses. For galaxies identified as post-starburst (E+A), the rate may be enhanced by an order of magnitude \cite{2016ApJ...818L..21F}, though we did not include such boosts in our fiducial calculations.

\subsection{Core-collapse and collapsar rates}

For comparison with other stellar transients, we also estimated baseline rates for core-collapse supernovae (CCSNe) and collapsars (long gamma-ray burst progenitors).
The CCSN rate in each galaxy was taken to scale directly with its star-formation rate (SFR) as
\begin{equation}
R_{\rm CCSN} = 0.01\,\mathrm{yr^{-1}}
\left(\frac{\mathrm{SFR}}{1,M_\odot,\mathrm{yr^{-1}}}\right),
\end{equation}
consistent with the empirical conversion adopted by \citet{Horiuchi2011CCSNR}.
Collapsar rates were estimated as a small fraction of the CCSN rate,
\begin{equation}
R_{\rm coll} = f_{\rm coll}R_{\rm CCSN},
\end{equation}
where we adopted a fiducial $f_{\rm coll}=10^{-3}$ following population studies of long GRBs and massive stellar core-collapse channels \citep{Woosley2006LGRB}.

\section{Fractional contribution at 60 EeV}

For completeness, we show in Table~\ref{tab:top10_new_60} the top ten contributing galaxies at 60\,EeV. Compared to the 32\,EeV case discussed in the main text, the dominance of the nearest sources becomes even more pronounced at higher energies, as the contribution from the rest of the Universe is increasingly suppressed.

\begin{table}[t]
\centering
\caption{Same as Table \ref{tab:top10}, but for $>60$\,EeV. Notice that the top 10 highest flux galaxies are not the same in this case, with M32 traded place with M106.}
\label{tab:top10_new_60}
\begin{tabular}{lcccccc}
\hline\hline
Galaxy       & Dist. & Allsky & Auger & TA & Hotspot & $E$ \\
             & [Mpc] & [\%]   & [\%]  & [\%] &        & [EeV] \\
\hline
Andromeda       & 0.8  & 24 & 0 & 37 & TA-PPSC    & $>25$ \\
M81 group       & 3.6  & 10 & 0 & 16 & TA-Ursa    & $>57$ \\
NGC253          & 3.5  &  4 & 12 & 0 &            &       \\
Cen~A           & 3.5  &  2 &  8 & 0 & Auger      & $>38$ \\
NGC4945         & 3.6  &  2 &  7 & 0 & Auger      & $>38$ \\
M33             & 0.9  &  2 &  0 & 3 & TA-PPSC    & $>25$ \\
IC342           & 3.4  &  2 &  0 & 3 &            &       \\
ESO097-013      & 4.2  &  1 &  4 & 0 & Auger      & $>38$ \\
M87             & 16.4 &  1 &  1 & 1 &            &       \\
M106         & 7.3  &  1 &  0 & 1 &            &       \\
\hline
Combined top-10 &      & 49 & 31 & 61 &            &       \\
$d>100$\,Mpc    &      &  5 &  4 &  2 &            &       \\
\hline\hline
\end{tabular}
\end{table}

\bibliographystyle{apsrev4-2}
\bibliography{refs} 

\begin{thebibliography}{62}%
\makeatletter
\providecommand \@ifxundefined [1]{%
 \@ifx{#1\undefined}
}%
\providecommand \@ifnum [1]{%
 \ifnum #1\expandafter \@firstoftwo
 \else \expandafter \@secondoftwo
 \fi
}%
\providecommand \@ifx [1]{%
 \ifx #1\expandafter \@firstoftwo
 \else \expandafter \@secondoftwo
 \fi
}%
\providecommand \natexlab [1]{#1}%
\providecommand \enquote  [1]{``#1''}%
\providecommand \bibnamefont  [1]{#1}%
\providecommand \bibfnamefont [1]{#1}%
\providecommand \citenamefont [1]{#1}%
\providecommand \href@noop [0]{\@secondoftwo}%
\providecommand \href [0]{\begingroup \@sanitize@url \@href}%
\providecommand \@href[1]{\@@startlink{#1}\@@href}%
\providecommand \@@href[1]{\endgroup#1\@@endlink}%
\providecommand \@sanitize@url [0]{\catcode `\\12\catcode `\$12\catcode `\&12\catcode `\#12\catcode `\^12\catcode `\_12\catcode `\%12\relax}%
\providecommand \@@startlink[1]{}%
\providecommand \@@endlink[0]{}%
\providecommand \url  [0]{\begingroup\@sanitize@url \@url }%
\providecommand \@url [1]{\endgroup\@href {#1}{\urlprefix }}%
\providecommand \urlprefix  [0]{URL }%
\providecommand \Eprint [0]{\href }%
\providecommand \doibase [0]{https://doi.org/}%
\providecommand \selectlanguage [0]{\@gobble}%
\providecommand \bibinfo  [0]{\@secondoftwo}%
\providecommand \bibfield  [0]{\@secondoftwo}%
\providecommand \translation [1]{[#1]}%
\providecommand \BibitemOpen [0]{}%
\providecommand \bibitemStop [0]{}%
\providecommand \bibitemNoStop [0]{.\EOS\space}%
\providecommand \EOS [0]{\spacefactor3000\relax}%
\providecommand \BibitemShut  [1]{\csname bibitem#1\endcsname}%
\let\auto@bib@innerbib\@empty
\bibitem [{\citenamefont {{Kotera}}\ and\ \citenamefont {{Olinto}}(2011)}]{2011ARA&A..49..119K}%
  \BibitemOpen
  \bibfield  {author} {\bibinfo {author} {\bibfnamefont {K.}~\bibnamefont {{Kotera}}}\ and\ \bibinfo {author} {\bibfnamefont {A.~V.}\ \bibnamefont {{Olinto}}},\ }\href {https://doi.org/10.1146/annurev-astro-081710-102620} {\bibfield  {journal} {\bibinfo  {journal} {\araa}\ }\textbf {\bibinfo {volume} {49}},\ \bibinfo {pages} {119} (\bibinfo {year} {2011})}\BibitemShut {NoStop}%
\bibitem [{\citenamefont {{Blasi}}(2013)}]{blasi2013}%
  \BibitemOpen
  \bibfield  {author} {\bibinfo {author} {\bibfnamefont {P.}~\bibnamefont {{Blasi}}},\ }\href {https://doi.org/10.1007/s00159-013-0070-7} {\bibfield  {journal} {\bibinfo  {journal} {\aapr}\ }\textbf {\bibinfo {volume} {21}},\ \bibinfo {eid} {70} (\bibinfo {year} {2013})}\BibitemShut {NoStop}%
\bibitem [{\citenamefont {Greisen}(1966)}]{greisen1966}%
  \BibitemOpen
  \bibfield  {author} {\bibinfo {author} {\bibfnamefont {K.}~\bibnamefont {Greisen}},\ }\href@noop {} {\bibfield  {journal} {\bibinfo  {journal} {\prl}\ }\textbf {\bibinfo {volume} {16}},\ \bibinfo {pages} {748} (\bibinfo {year} {1966})}\BibitemShut {NoStop}%
\bibitem [{\citenamefont {Zatsepin}\ and\ \citenamefont {Kuzmin}(1966)}]{zatsepin1966}%
  \BibitemOpen
  \bibfield  {author} {\bibinfo {author} {\bibfnamefont {G.}~\bibnamefont {Zatsepin}}\ and\ \bibinfo {author} {\bibfnamefont {V.}~\bibnamefont {Kuzmin}},\ }\href@noop {} {\bibfield  {journal} {\bibinfo  {journal} {JETP Lett.}\ }\textbf {\bibinfo {volume} {4}},\ \bibinfo {pages} {78} (\bibinfo {year} {1966})}\BibitemShut {NoStop}%
\bibitem [{\citenamefont {{Allard}}(2012)}]{2012APh....39...33A}%
  \BibitemOpen
  \bibfield  {author} {\bibinfo {author} {\bibfnamefont {D.}~\bibnamefont {{Allard}}},\ }\href {https://doi.org/10.1016/j.astropartphys.2011.10.011} {\bibfield  {journal} {\bibinfo  {journal} {\apph}\ }\textbf {\bibinfo {volume} {39}},\ \bibinfo {pages} {33} (\bibinfo {year} {2012})}\BibitemShut {NoStop}%
\bibitem [{\citenamefont {{van Vliet}}\ \emph {et~al.}(2022)\citenamefont {{van Vliet}} \emph {et~al.}}]{2022MNRAS.510.1289V}%
  \BibitemOpen
  \bibfield  {author} {\bibinfo {author} {\bibfnamefont {A.}~\bibnamefont {{van Vliet}}} \emph {et~al.},\ }\href {https://doi.org/10.1093/mnras/stab3495} {\bibfield  {journal} {\bibinfo  {journal} {\mnras}\ }\textbf {\bibinfo {volume} {510}},\ \bibinfo {pages} {1289} (\bibinfo {year} {2022})}\BibitemShut {NoStop}%
\bibitem [{\citenamefont {Aab}\ \emph {et~al.}(2015)\citenamefont {Aab} \emph {et~al.}}]{2015172}%
  \BibitemOpen
  \bibfield  {author} {\bibinfo {author} {\bibfnamefont {A.}~\bibnamefont {Aab}} \emph {et~al.},\ }\href {https://doi.org/https://doi.org/10.1016/j.nima.2015.06.058} {\bibfield  {journal} {\bibinfo  {journal} {Nucl. Instrum. Meth. A}\ }\textbf {\bibinfo {volume} {798}},\ \bibinfo {pages} {172} (\bibinfo {year} {2015})}\BibitemShut {NoStop}%
\bibitem [{\citenamefont {{Abreu}}\ \emph {et~al.}(2022)\citenamefont {{Abreu}} \emph {et~al.}}]{2022ApJ...935..170A}%
  \BibitemOpen
  \bibfield  {author} {\bibinfo {author} {\bibfnamefont {P.}~\bibnamefont {{Abreu}}} \emph {et~al.},\ }\href {https://doi.org/10.3847/1538-4357/ac7d4e} {\bibfield  {journal} {\bibinfo  {journal} {\apj}\ }\textbf {\bibinfo {volume} {935}},\ \bibinfo {eid} {170} (\bibinfo {year} {2022})}\BibitemShut {NoStop}%
\bibitem [{\citenamefont {Abu-Zayyad}\ \emph {et~al.}(2012)\citenamefont {Abu-Zayyad} \emph {et~al.}}]{ABUZAYYAD201287}%
  \BibitemOpen
  \bibfield  {author} {\bibinfo {author} {\bibfnamefont {T.}~\bibnamefont {Abu-Zayyad}} \emph {et~al.},\ }\href {https://doi.org/https://doi.org/10.1016/j.nima.2012.05.079} {\bibfield  {journal} {\bibinfo  {journal} {Nucl. Instrum. Meth. A}\ }\textbf {\bibinfo {volume} {689}},\ \bibinfo {pages} {87} (\bibinfo {year} {2012})}\BibitemShut {NoStop}%
\bibitem [{\citenamefont {{Abbasi}}\ \emph {et~al.}(2021)\citenamefont {{Abbasi}} \emph {et~al.}}]{TAhotspot25}%
  \BibitemOpen
  \bibfield  {author} {\bibinfo {author} {\bibfnamefont {R.~U.}\ \bibnamefont {{Abbasi}}} \emph {et~al.},\ }\href@noop {} {\bibfield  {journal} {\bibinfo  {journal} {arXiv:2110.14827}\ } (\bibinfo {year} {2021})}\BibitemShut {NoStop}%
\bibitem [{\citenamefont {{Abbasi}}\ \emph {et~al.}(2014)\citenamefont {{Abbasi}} \emph {et~al.}}]{ta2014hotspot57}%
  \BibitemOpen
  \bibfield  {author} {\bibinfo {author} {\bibfnamefont {R.~U.}\ \bibnamefont {{Abbasi}}} \emph {et~al.},\ }\href {https://doi.org/10.1088/2041-8205/790/2/L21} {\bibfield  {journal} {\bibinfo  {journal} {\apjl}\ }\textbf {\bibinfo {volume} {790}},\ \bibinfo {eid} {L21} (\bibinfo {year} {2014})}\BibitemShut {NoStop}%
\bibitem [{\citenamefont {{Aab}}\ \emph {et~al.}(2017{\natexlab{a}})\citenamefont {{Aab}} \emph {et~al.}}]{Auger2017dipole}%
  \BibitemOpen
  \bibfield  {author} {\bibinfo {author} {\bibfnamefont {A.}~\bibnamefont {{Aab}}} \emph {et~al.},\ }\href {https://doi.org/10.1126/science.aan4338} {\bibfield  {journal} {\bibinfo  {journal} {Science}\ }\textbf {\bibinfo {volume} {357}},\ \bibinfo {pages} {1266} (\bibinfo {year} {2017}{\natexlab{a}})}\BibitemShut {NoStop}%
\bibitem [{\citenamefont {{Bister}}\ \emph {et~al.}(2024)\citenamefont {{Bister}}, \citenamefont {{Farrar}},\ and\ \citenamefont {{Unger}}}]{2024ApJ...975L..21B}%
  \BibitemOpen
  \bibfield  {author} {\bibinfo {author} {\bibfnamefont {T.}~\bibnamefont {{Bister}}}, \bibinfo {author} {\bibfnamefont {G.~R.}\ \bibnamefont {{Farrar}}},\ and\ \bibinfo {author} {\bibfnamefont {M.}~\bibnamefont {{Unger}}},\ }\href {https://doi.org/10.3847/2041-8213/ad856f} {\bibfield  {journal} {\bibinfo  {journal} {\apjl}\ }\textbf {\bibinfo {volume} {975}},\ \bibinfo {eid} {L21} (\bibinfo {year} {2024})},\ \Eprint {https://arxiv.org/abs/2408.00614} {arXiv:2408.00614 [astro-ph.HE]} \BibitemShut {NoStop}%
\bibitem [{\citenamefont {{Aab}}\ \emph {et~al.}(2017{\natexlab{b}})\citenamefont {{Aab}} \emph {et~al.}}]{auger2017composition}%
  \BibitemOpen
  \bibfield  {author} {\bibinfo {author} {\bibfnamefont {A.}~\bibnamefont {{Aab}}} \emph {et~al.},\ }\href {https://doi.org/10.1088/1475-7516/2017/04/038} {\bibfield  {journal} {\bibinfo  {journal} {\jcap}\ }\textbf {\bibinfo {volume} {2017}},\ \bibinfo {eid} {038} (\bibinfo {year} {2017}{\natexlab{b}})}\BibitemShut {NoStop}%
\bibitem [{\citenamefont {{Aab}}\ \emph {et~al.}(2020)\citenamefont {{Aab}} \emph {et~al.}}]{Auger2020PRDSpectrum}%
  \BibitemOpen
  \bibfield  {author} {\bibinfo {author} {\bibfnamefont {A.}~\bibnamefont {{Aab}}} \emph {et~al.},\ }\href {https://doi.org/10.1103/PhysRevD.102.062005} {\bibfield  {journal} {\bibinfo  {journal} {\prd}\ }\textbf {\bibinfo {volume} {102}},\ \bibinfo {eid} {062005} (\bibinfo {year} {2020})}\BibitemShut {NoStop}%
\bibitem [{\citenamefont {{Abbasi}}\ \emph {et~al.}(2023)\citenamefont {{Abbasi}} \emph {et~al.}}]{TAspectrum2023}%
  \BibitemOpen
  \bibfield  {author} {\bibinfo {author} {\bibfnamefont {R.~U.}\ \bibnamefont {{Abbasi}}} \emph {et~al.},\ }\href {https://doi.org/10.1016/j.astropartphys.2023.102864} {\bibfield  {journal} {\bibinfo  {journal} {\apph}\ }\textbf {\bibinfo {volume} {151}},\ \bibinfo {eid} {102864} (\bibinfo {year} {2023})}\BibitemShut {NoStop}%
\bibitem [{\citenamefont {{Plotko}}\ \emph {et~al.}(2023)\citenamefont {{Plotko}} \emph {et~al.}}]{Plotko:2022urd}%
  \BibitemOpen
  \bibfield  {author} {\bibinfo {author} {\bibfnamefont {P.}~\bibnamefont {{Plotko}}} \emph {et~al.},\ }\href {https://doi.org/10.3847/1538-4357/acdf59} {\bibfield  {journal} {\bibinfo  {journal} {\apj}\ }\textbf {\bibinfo {volume} {953}},\ \bibinfo {eid} {129} (\bibinfo {year} {2023})}\BibitemShut {NoStop}%
\bibitem [{\citenamefont {{Abraham}}\ \emph {et~al.}(2007)\citenamefont {{Abraham}} \emph {et~al.}}]{abraham2007agn}%
  \BibitemOpen
  \bibfield  {author} {\bibinfo {author} {\bibfnamefont {J.}~\bibnamefont {{Abraham}}} \emph {et~al.},\ }\href {https://doi.org/10.1126/science.1151124} {\bibfield  {journal} {\bibinfo  {journal} {Science}\ }\textbf {\bibinfo {volume} {318}},\ \bibinfo {pages} {938} (\bibinfo {year} {2007})}\BibitemShut {NoStop}%
\bibitem [{\citenamefont {{Aab}}\ \emph {et~al.}(2015)\citenamefont {{Aab}} \emph {et~al.}}]{2015ApJ...804...15A}%
  \BibitemOpen
  \bibfield  {author} {\bibinfo {author} {\bibfnamefont {A.}~\bibnamefont {{Aab}}} \emph {et~al.},\ }\href {https://doi.org/10.1088/0004-637X/804/1/15} {\bibfield  {journal} {\bibinfo  {journal} {\apj}\ }\textbf {\bibinfo {volume} {804}},\ \bibinfo {eid} {15} (\bibinfo {year} {2015})}\BibitemShut {NoStop}%
\bibitem [{\citenamefont {{Waxman}}(1995)}]{Waxman:1995vg}%
  \BibitemOpen
  \bibfield  {author} {\bibinfo {author} {\bibfnamefont {E.}~\bibnamefont {{Waxman}}},\ }\href {https://doi.org/10.1103/PhysRevLett.75.386} {\bibfield  {journal} {\bibinfo  {journal} {\prl}\ }\textbf {\bibinfo {volume} {75}},\ \bibinfo {pages} {386} (\bibinfo {year} {1995})}\BibitemShut {NoStop}%
\bibitem [{\citenamefont {Anchordoqui}(2019)}]{anchordoqui2019review}%
  \BibitemOpen
  \bibfield  {author} {\bibinfo {author} {\bibfnamefont {L.~A.}\ \bibnamefont {Anchordoqui}},\ }\href {https://doi.org/https://doi.org/10.1016/j.physrep.2019.01.002} {\bibfield  {journal} {\bibinfo  {journal} {Phys.\ Rep.}\ }\textbf {\bibinfo {volume} {801}},\ \bibinfo {pages} {1} (\bibinfo {year} {2019})},\ \bibinfo {note} {ultra-high-energy cosmic rays}\BibitemShut {NoStop}%
\bibitem [{\citenamefont {Aab}\ \emph {et~al.}(2018)\citenamefont {Aab}, \citenamefont {Abreu}, \citenamefont {Aglietta}, \citenamefont {Albuquerque}, \citenamefont {Allekotte}, \citenamefont {Almela} \emph {et~al.}}]{Aab_2018}%
  \BibitemOpen
  \bibfield  {author} {\bibinfo {author} {\bibfnamefont {A.}~\bibnamefont {Aab}}, \bibinfo {author} {\bibfnamefont {P.}~\bibnamefont {Abreu}}, \bibinfo {author} {\bibfnamefont {M.}~\bibnamefont {Aglietta}}, \bibinfo {author} {\bibfnamefont {I.~F.~M.}\ \bibnamefont {Albuquerque}}, \bibinfo {author} {\bibfnamefont {I.}~\bibnamefont {Allekotte}}, \bibinfo {author} {\bibfnamefont {A.}~\bibnamefont {Almela}}, \emph {et~al.},\ }\href {https://doi.org/10.3847/2041-8213/aaa66d} {\bibfield  {journal} {\bibinfo  {journal} {\apjl}\ }\textbf {\bibinfo {volume} {853}},\ \bibinfo {pages} {L29} (\bibinfo {year} {2018})}\BibitemShut {NoStop}%
\bibitem [{\citenamefont {{Waxman}}\ and\ \citenamefont {{Miralda-Escude}}(1996)}]{Waxman:1996zn}%
  \BibitemOpen
  \bibfield  {author} {\bibinfo {author} {\bibfnamefont {E.}~\bibnamefont {{Waxman}}}\ and\ \bibinfo {author} {\bibfnamefont {J.}~\bibnamefont {{Miralda-Escude}}},\ }\href {https://doi.org/10.1086/310367} {\bibfield  {journal} {\bibinfo  {journal} {\apjl}\ }\textbf {\bibinfo {volume} {472}},\ \bibinfo {pages} {L89} (\bibinfo {year} {1996})}\BibitemShut {NoStop}%
\bibitem [{\citenamefont {{Harari}}\ \emph {et~al.}(2000)\citenamefont {{Harari}}, \citenamefont {{Mollerach}},\ and\ \citenamefont {{Roulet}}}]{2000JHEP...02..035H}%
  \BibitemOpen
  \bibfield  {author} {\bibinfo {author} {\bibfnamefont {D.}~\bibnamefont {{Harari}}}, \bibinfo {author} {\bibfnamefont {S.}~\bibnamefont {{Mollerach}}},\ and\ \bibinfo {author} {\bibfnamefont {E.}~\bibnamefont {{Roulet}}},\ }\href {https://doi.org/10.1088/1126-6708/2000/02/035} {\bibfield  {journal} {\bibinfo  {journal} {J.\ High\ Energy\ Phys.}\ }\textbf {\bibinfo {volume} {2000}},\ \bibinfo {eid} {035} (\bibinfo {year} {2000})}\BibitemShut {NoStop}%
\bibitem [{\citenamefont {{Taylor}}\ \emph {et~al.}(2023)\citenamefont {{Taylor}}, \citenamefont {{Matthews}},\ and\ \citenamefont {{Bell}}}]{2023MNRAS.524..631T}%
  \BibitemOpen
  \bibfield  {author} {\bibinfo {author} {\bibfnamefont {A.~M.}\ \bibnamefont {{Taylor}}}, \bibinfo {author} {\bibfnamefont {J.~H.}\ \bibnamefont {{Matthews}}},\ and\ \bibinfo {author} {\bibfnamefont {A.~R.}\ \bibnamefont {{Bell}}},\ }\href {https://doi.org/10.1093/mnras/stad1716} {\bibfield  {journal} {\bibinfo  {journal} {\mnras}\ }\textbf {\bibinfo {volume} {524}},\ \bibinfo {pages} {631} (\bibinfo {year} {2023})}\BibitemShut {NoStop}%
\bibitem [{\citenamefont {{Bister}}\ and\ \citenamefont {{Farrar}}(2024)}]{2024ApJ...966...71B}%
  \BibitemOpen
  \bibfield  {author} {\bibinfo {author} {\bibfnamefont {T.}~\bibnamefont {{Bister}}}\ and\ \bibinfo {author} {\bibfnamefont {G.~R.}\ \bibnamefont {{Farrar}}},\ }\href {https://doi.org/10.3847/1538-4357/ad2f3f} {\bibfield  {journal} {\bibinfo  {journal} {\apj}\ }\textbf {\bibinfo {volume} {966}},\ \bibinfo {eid} {71} (\bibinfo {year} {2024})},\ \Eprint {https://arxiv.org/abs/2312.02645} {arXiv:2312.02645 [astro-ph.HE]} \BibitemShut {NoStop}%
\bibitem [{\citenamefont {{Marafico}}\ \emph {et~al.}(2024)\citenamefont {{Marafico}}, \citenamefont {{Biteau}}, \citenamefont {{Condorelli}}, \citenamefont {{Deligny}},\ and\ \citenamefont {{Bregeon}}}]{2024ApJ...972....4M}%
  \BibitemOpen
  \bibfield  {author} {\bibinfo {author} {\bibfnamefont {S.}~\bibnamefont {{Marafico}}}, \bibinfo {author} {\bibfnamefont {J.}~\bibnamefont {{Biteau}}}, \bibinfo {author} {\bibfnamefont {A.}~\bibnamefont {{Condorelli}}}, \bibinfo {author} {\bibfnamefont {O.}~\bibnamefont {{Deligny}}},\ and\ \bibinfo {author} {\bibfnamefont {J.}~\bibnamefont {{Bregeon}}},\ }\href {https://doi.org/10.3847/1538-4357/ad5a11} {\bibfield  {journal} {\bibinfo  {journal} {\apj}\ }\textbf {\bibinfo {volume} {972}},\ \bibinfo {eid} {4} (\bibinfo {year} {2024})},\ \Eprint {https://arxiv.org/abs/2405.17179} {arXiv:2405.17179 [astro-ph.HE]} \BibitemShut {NoStop}%
\bibitem [{\citenamefont {{Unger}}\ and\ \citenamefont {{Farrar}}(2024)}]{2024ApJ...962L...5U}%
  \BibitemOpen
  \bibfield  {author} {\bibinfo {author} {\bibfnamefont {M.}~\bibnamefont {{Unger}}}\ and\ \bibinfo {author} {\bibfnamefont {G.~R.}\ \bibnamefont {{Farrar}}},\ }\href {https://doi.org/10.3847/2041-8213/ad1ced} {\bibfield  {journal} {\bibinfo  {journal} {\apjl}\ }\textbf {\bibinfo {volume} {962}},\ \bibinfo {eid} {L5} (\bibinfo {year} {2024})},\ \Eprint {https://arxiv.org/abs/2312.13273} {arXiv:2312.13273 [astro-ph.HE]} \BibitemShut {NoStop}%
\bibitem [{\citenamefont {{Takami}}\ \emph {et~al.}(2014)\citenamefont {{Takami}}, \citenamefont {{Kyutoku}},\ and\ \citenamefont {{Ioka}}}]{2014PhRvD..89f3006T}%
  \BibitemOpen
  \bibfield  {author} {\bibinfo {author} {\bibfnamefont {H.}~\bibnamefont {{Takami}}}, \bibinfo {author} {\bibfnamefont {K.}~\bibnamefont {{Kyutoku}}},\ and\ \bibinfo {author} {\bibfnamefont {K.}~\bibnamefont {{Ioka}}},\ }\href {https://doi.org/10.1103/PhysRevD.89.063006} {\bibfield  {journal} {\bibinfo  {journal} {\prd}\ }\textbf {\bibinfo {volume} {89}},\ \bibinfo {eid} {063006} (\bibinfo {year} {2014})},\ \Eprint {https://arxiv.org/abs/1307.6805} {arXiv:1307.6805 [astro-ph.HE]} \BibitemShut {NoStop}%
\bibitem [{\citenamefont {{Kimura}}\ \emph {et~al.}(2018)\citenamefont {{Kimura}}, \citenamefont {{Murase}},\ and\ \citenamefont {{M{\'e}sz{\'a}ros}}}]{2018ApJ...866...51K}%
  \BibitemOpen
  \bibfield  {author} {\bibinfo {author} {\bibfnamefont {S.~S.}\ \bibnamefont {{Kimura}}}, \bibinfo {author} {\bibfnamefont {K.}~\bibnamefont {{Murase}}},\ and\ \bibinfo {author} {\bibfnamefont {P.}~\bibnamefont {{M{\'e}sz{\'a}ros}}},\ }\href {https://doi.org/10.3847/1538-4357/aadc0a} {\bibfield  {journal} {\bibinfo  {journal} {\apj}\ }\textbf {\bibinfo {volume} {866}},\ \bibinfo {eid} {51} (\bibinfo {year} {2018})},\ \Eprint {https://arxiv.org/abs/1807.03290} {arXiv:1807.03290 [astro-ph.HE]} \BibitemShut {NoStop}%
\bibitem [{\citenamefont {{Zhang}}\ \emph {et~al.}(2024)\citenamefont {{Zhang}}, \citenamefont {{Murase}}, \citenamefont {{Ekanger}}, \citenamefont {{Bhattacharya}},\ and\ \citenamefont {{Horiuchi}}}]{2024arXiv240517409Z}%
  \BibitemOpen
  \bibfield  {author} {\bibinfo {author} {\bibfnamefont {B.~T.}\ \bibnamefont {{Zhang}}}, \bibinfo {author} {\bibfnamefont {K.}~\bibnamefont {{Murase}}}, \bibinfo {author} {\bibfnamefont {N.}~\bibnamefont {{Ekanger}}}, \bibinfo {author} {\bibfnamefont {M.}~\bibnamefont {{Bhattacharya}}},\ and\ \bibinfo {author} {\bibfnamefont {S.}~\bibnamefont {{Horiuchi}}},\ }\href {https://doi.org/10.48550/arXiv.2405.17409} {\bibfield  {journal} {\bibinfo  {journal} {arXiv e-prints}\ ,\ \bibinfo {eid} {arXiv:2405.17409}} (\bibinfo {year} {2024})},\ \Eprint {https://arxiv.org/abs/2405.17409} {arXiv:2405.17409 [astro-ph.HE]} \BibitemShut {NoStop}%
\bibitem [{\citenamefont {{Farrar}}(2025{\natexlab{a}})}]{2025PhRvL.134h1003F}%
  \BibitemOpen
  \bibfield  {author} {\bibinfo {author} {\bibfnamefont {G.~R.}\ \bibnamefont {{Farrar}}},\ }\href {https://doi.org/10.1103/PhysRevLett.134.081003} {\bibfield  {journal} {\bibinfo  {journal} {\prl}\ }\textbf {\bibinfo {volume} {134}},\ \bibinfo {eid} {081003} (\bibinfo {year} {2025}{\natexlab{a}})}\BibitemShut {NoStop}%
\bibitem [{\citenamefont {{Farrar}}(2025{\natexlab{b}})}]{2025arXiv250622625F}%
  \BibitemOpen
  \bibfield  {author} {\bibinfo {author} {\bibfnamefont {G.~R.}\ \bibnamefont {{Farrar}}},\ }\href@noop {} {\bibfield  {journal} {\bibinfo  {journal} {arXiv:2506.22625}\ } (\bibinfo {year} {2025}{\natexlab{b}})}\BibitemShut {NoStop}%
\bibitem [{\citenamefont {{Farrar}}\ and\ \citenamefont {{Piran}}(2014)}]{2014arXiv1411.0704F}%
  \BibitemOpen
  \bibfield  {author} {\bibinfo {author} {\bibfnamefont {G.~R.}\ \bibnamefont {{Farrar}}}\ and\ \bibinfo {author} {\bibfnamefont {T.}~\bibnamefont {{Piran}}},\ }\href@noop {} {\bibfield  {journal} {\bibinfo  {journal} {arXiv:1411.0704}\ } (\bibinfo {year} {2014})}\BibitemShut {NoStop}%
\bibitem [{\citenamefont {Zhang}\ \emph {et~al.}(2017)\citenamefont {Zhang} \emph {et~al.}}]{PhysRevD.96.063007}%
  \BibitemOpen
  \bibfield  {author} {\bibinfo {author} {\bibfnamefont {B.~T.}\ \bibnamefont {Zhang}} \emph {et~al.},\ }\href {https://doi.org/10.1103/PhysRevD.96.063007} {\bibfield  {journal} {\bibinfo  {journal} {\prd}\ }\textbf {\bibinfo {volume} {96}},\ \bibinfo {pages} {063007} (\bibinfo {year} {2017})}\BibitemShut {NoStop}%
\bibitem [{\citenamefont {{Kovlakas}}\ \emph {et~al.}(2021)\citenamefont {{Kovlakas}} \emph {et~al.}}]{Kovlakas2021HECATE}%
  \BibitemOpen
  \bibfield  {author} {\bibinfo {author} {\bibfnamefont {K.}~\bibnamefont {{Kovlakas}}} \emph {et~al.},\ }\href {https://doi.org/10.1093/mnras/stab1799} {\bibfield  {journal} {\bibinfo  {journal} {\mnras}\ }\textbf {\bibinfo {volume} {506}},\ \bibinfo {pages} {1896} (\bibinfo {year} {2021})}\BibitemShut {NoStop}%
\bibitem [{\citenamefont {{Tully}}\ \emph {et~al.}(2023)\citenamefont {{Tully}} \emph {et~al.}}]{CF4}%
  \BibitemOpen
  \bibfield  {author} {\bibinfo {author} {\bibfnamefont {R.~B.}\ \bibnamefont {{Tully}}} \emph {et~al.},\ }\href {https://doi.org/10.3847/1538-4357/ac94d8} {\bibfield  {journal} {\bibinfo  {journal} {\apj}\ }\textbf {\bibinfo {volume} {944}},\ \bibinfo {eid} {94} (\bibinfo {year} {2023})}\BibitemShut {NoStop}%
\bibitem [{\citenamefont {{Karachentsev}}\ \emph {et~al.}(2013)\citenamefont {{Karachentsev}}, \citenamefont {{Makarov}},\ and\ \citenamefont {{Kaisina}}}]{UNGC}%
  \BibitemOpen
  \bibfield  {author} {\bibinfo {author} {\bibfnamefont {I.~D.}\ \bibnamefont {{Karachentsev}}}, \bibinfo {author} {\bibfnamefont {D.~I.}\ \bibnamefont {{Makarov}}},\ and\ \bibinfo {author} {\bibfnamefont {E.~I.}\ \bibnamefont {{Kaisina}}},\ }\href {https://doi.org/10.1088/0004-6256/145/4/101} {\bibfield  {journal} {\bibinfo  {journal} {\aj}\ }\textbf {\bibinfo {volume} {145}},\ \bibinfo {eid} {101} (\bibinfo {year} {2013})}\BibitemShut {NoStop}%
\bibitem [{\citenamefont {{Cutri}}\ \emph {et~al.}(2013)\citenamefont {{Cutri}} \emph {et~al.}}]{allwise}%
  \BibitemOpen
  \bibfield  {author} {\bibinfo {author} {\bibfnamefont {R.~M.}\ \bibnamefont {{Cutri}}} \emph {et~al.},\ }\href@noop {} {\bibinfo {title} {{Explanatory Supplement to the AllWISE Data Release Products}}} (\bibinfo {year} {2013})\BibitemShut {NoStop}%
\bibitem [{\citenamefont {{Skrutskie}}\ \emph {et~al.}(2006)\citenamefont {{Skrutskie}} \emph {et~al.}}]{Skrutskie2006TwoMASS}%
  \BibitemOpen
  \bibfield  {author} {\bibinfo {author} {\bibfnamefont {M.~F.}\ \bibnamefont {{Skrutskie}}} \emph {et~al.},\ }\href {https://doi.org/10.1086/498708} {\bibfield  {journal} {\bibinfo  {journal} {\aj}\ }\textbf {\bibinfo {volume} {131}},\ \bibinfo {pages} {1163} (\bibinfo {year} {2006})}\BibitemShut {NoStop}%
\bibitem [{\citenamefont {{Jarrett}}\ \emph {et~al.}(2000)\citenamefont {{Jarrett}} \emph {et~al.}}]{Jarrett2000XSC}%
  \BibitemOpen
  \bibfield  {author} {\bibinfo {author} {\bibfnamefont {T.~H.}\ \bibnamefont {{Jarrett}}} \emph {et~al.},\ }\href {https://doi.org/10.1086/301330} {\bibfield  {journal} {\bibinfo  {journal} {\aj}\ }\textbf {\bibinfo {volume} {119}},\ \bibinfo {pages} {2498} (\bibinfo {year} {2000})}\BibitemShut {NoStop}%
\bibitem [{\citenamefont {{Cluver}}\ \emph {et~al.}(2014)\citenamefont {{Cluver}} \emph {et~al.}}]{Cluver2014WISEML}%
  \BibitemOpen
  \bibfield  {author} {\bibinfo {author} {\bibfnamefont {M.~E.}\ \bibnamefont {{Cluver}}} \emph {et~al.},\ }\href {https://doi.org/10.1088/0004-637X/782/2/90} {\bibfield  {journal} {\bibinfo  {journal} {\apj}\ }\textbf {\bibinfo {volume} {782}},\ \bibinfo {eid} {90} (\bibinfo {year} {2014})}\BibitemShut {NoStop}%
\bibitem [{\citenamefont {{Willmer}}(2018)}]{Willmer2018SolarMags}%
  \BibitemOpen
  \bibfield  {author} {\bibinfo {author} {\bibfnamefont {C.~N.~A.}\ \bibnamefont {{Willmer}}},\ }\href {https://doi.org/10.3847/1538-4365/aabfdf} {\bibfield  {journal} {\bibinfo  {journal} {\apjs}\ }\textbf {\bibinfo {volume} {236}},\ \bibinfo {eid} {47} (\bibinfo {year} {2018})}\BibitemShut {NoStop}%
\bibitem [{\citenamefont {{Artale}}\ \emph {et~al.}(2020)\citenamefont {{Artale}}, \citenamefont {{Mapelli}}, \citenamefont {{Bouffanais}}, \citenamefont {{Giacobbo}}, \citenamefont {{Pasquato}},\ and\ \citenamefont {{Spera}}}]{2020MNRAS.491.3419A}%
  \BibitemOpen
  \bibfield  {author} {\bibinfo {author} {\bibfnamefont {M.~C.}\ \bibnamefont {{Artale}}}, \bibinfo {author} {\bibfnamefont {M.}~\bibnamefont {{Mapelli}}}, \bibinfo {author} {\bibfnamefont {Y.}~\bibnamefont {{Bouffanais}}}, \bibinfo {author} {\bibfnamefont {N.}~\bibnamefont {{Giacobbo}}}, \bibinfo {author} {\bibfnamefont {M.}~\bibnamefont {{Pasquato}}},\ and\ \bibinfo {author} {\bibfnamefont {M.}~\bibnamefont {{Spera}}},\ }\href {https://doi.org/10.1093/mnras/stz3190} {\bibfield  {journal} {\bibinfo  {journal} {\mnras}\ }\textbf {\bibinfo {volume} {491}},\ \bibinfo {pages} {3419} (\bibinfo {year} {2020})}\BibitemShut {NoStop}%
\bibitem [{\citenamefont {{Abac}}\ \emph {et~al.}(2025)\citenamefont {{Abac}} \emph {et~al.}}]{2025arXiv250818083T}%
  \BibitemOpen
  \bibfield  {author} {\bibinfo {author} {\bibfnamefont {A.~G.}\ \bibnamefont {{Abac}}} \emph {et~al.},\ }\href@noop {} {\bibfield  {journal} {\bibinfo  {journal} {arXiv:2508.18083}\ } (\bibinfo {year} {2025})}\BibitemShut {NoStop}%
\bibitem [{\citenamefont {{Aab}}\ \emph {et~al.}(2017{\natexlab{c}})\citenamefont {{Aab}}, \citenamefont {{Abreu}}, \citenamefont {{Aglietta}}, \citenamefont {{Samarai}} \emph {et~al.}}]{2017JCAP...04..038A}%
  \BibitemOpen
  \bibfield  {author} {\bibinfo {author} {\bibfnamefont {A.}~\bibnamefont {{Aab}}}, \bibinfo {author} {\bibfnamefont {P.}~\bibnamefont {{Abreu}}}, \bibinfo {author} {\bibfnamefont {M.}~\bibnamefont {{Aglietta}}}, \bibinfo {author} {\bibfnamefont {I.~A.}\ \bibnamefont {{Samarai}}}, \emph {et~al.},\ }\href {https://doi.org/10.1088/1475-7516/2017/04/038} {\bibfield  {journal} {\bibinfo  {journal} {\jcap}\ }\textbf {\bibinfo {volume} {2017}},\ \bibinfo {eid} {038} (\bibinfo {year} {2017}{\natexlab{c}})},\ \Eprint {https://arxiv.org/abs/1612.07155} {arXiv:1612.07155 [astro-ph.HE]} \BibitemShut {NoStop}%
\bibitem [{\citenamefont {Aloisio}\ \emph {et~al.}(2017)\citenamefont {Aloisio}, \citenamefont {Boncioli}, \citenamefont {di~Matteo}, \citenamefont {Grillo}, \citenamefont {Petrera},\ and\ \citenamefont {Salamida}}]{Aloisio_2017}%
  \BibitemOpen
  \bibfield  {author} {\bibinfo {author} {\bibfnamefont {R.}~\bibnamefont {Aloisio}}, \bibinfo {author} {\bibfnamefont {D.}~\bibnamefont {Boncioli}}, \bibinfo {author} {\bibfnamefont {A.}~\bibnamefont {di~Matteo}}, \bibinfo {author} {\bibfnamefont {A.~F.}\ \bibnamefont {Grillo}}, \bibinfo {author} {\bibfnamefont {S.}~\bibnamefont {Petrera}},\ and\ \bibinfo {author} {\bibfnamefont {F.}~\bibnamefont {Salamida}},\ }\href {https://doi.org/10.1088/1475-7516/2017/11/009} {\bibfield  {journal} {\bibinfo  {journal} {Journal of Cosmology and Astroparticle Physics}\ }\textbf {\bibinfo {volume} {2017}}\bibinfo  {number} { (11)},\ \bibinfo {pages} {009}}\BibitemShut {NoStop}%
\bibitem [{\citenamefont {{McCall}}(2014)}]{2014MNRAS.440..405M}%
  \BibitemOpen
\bibfield  {number} {  }\bibfield  {author} {\bibinfo {author} {\bibfnamefont {M.~L.}\ \bibnamefont {{McCall}}},\ }\href {https://doi.org/10.1093/mnras/stu199} {\bibfield  {journal} {\bibinfo  {journal} {\mnras}\ }\textbf {\bibinfo {volume} {440}},\ \bibinfo {pages} {405} (\bibinfo {year} {2014})},\ \Eprint {https://arxiv.org/abs/1403.3667} {arXiv:1403.3667 [astro-ph.GA]} \BibitemShut {NoStop}%
\bibitem [{\citenamefont {Neronov}\ \emph {et~al.}(2023)\citenamefont {Neronov}, \citenamefont {Semikoz},\ and\ \citenamefont {Kalashev}}]{PhysRevD.108.103008}%
  \BibitemOpen
  \bibfield  {author} {\bibinfo {author} {\bibfnamefont {A.}~\bibnamefont {Neronov}}, \bibinfo {author} {\bibfnamefont {D.}~\bibnamefont {Semikoz}},\ and\ \bibinfo {author} {\bibfnamefont {O.}~\bibnamefont {Kalashev}},\ }\href {https://doi.org/10.1103/PhysRevD.108.103008} {\bibfield  {journal} {\bibinfo  {journal} {Phys. Rev. D}\ }\textbf {\bibinfo {volume} {108}},\ \bibinfo {pages} {103008} (\bibinfo {year} {2023})}\BibitemShut {NoStop}%
\bibitem [{\citenamefont {{He}}\ \emph {et~al.}(2016)\citenamefont {{He}} \emph {et~al.}}]{2016PhRvD..93d3011H}%
  \BibitemOpen
  \bibfield  {author} {\bibinfo {author} {\bibfnamefont {H.-N.}\ \bibnamefont {{He}}} \emph {et~al.},\ }\href {https://doi.org/10.1103/PhysRevD.93.043011} {\bibfield  {journal} {\bibinfo  {journal} {\prd}\ }\textbf {\bibinfo {volume} {93}},\ \bibinfo {eid} {043011} (\bibinfo {year} {2016})}\BibitemShut {NoStop}%
\bibitem [{\citenamefont {{Sommers}}(2001)}]{Sommers2001}%
  \BibitemOpen
  \bibfield  {author} {\bibinfo {author} {\bibfnamefont {P.}~\bibnamefont {{Sommers}}},\ }\href {https://doi.org/10.1016/S0927-6505(00)00130-4} {\bibfield  {journal} {\bibinfo  {journal} {\apph}\ }\textbf {\bibinfo {volume} {14}},\ \bibinfo {pages} {271} (\bibinfo {year} {2001})}\BibitemShut {NoStop}%
\bibitem [{\citenamefont {Kim}\ \emph {et~al.}(2025)\citenamefont {Kim}, \citenamefont {Ivanov},\ and\ \citenamefont {Thomson}}]{Kim:2025qmo}%
  \BibitemOpen
  \bibfield  {author} {\bibinfo {author} {\bibfnamefont {J.}~\bibnamefont {Kim}}, \bibinfo {author} {\bibfnamefont {D.}~\bibnamefont {Ivanov}},\ and\ \bibinfo {author} {\bibfnamefont {G.}~\bibnamefont {Thomson}},\ }\href {https://doi.org/10.22323/1.501.0301} {\bibfield  {journal} {\bibinfo  {journal} {PoS}\ }\textbf {\bibinfo {volume} {ICRC2025}},\ \bibinfo {pages} {301} (\bibinfo {year} {2025})},\ \bibinfo {note} {conference presentation at the 39th International Cosmic Ray Conference (ICRC 2025), Geneva, Switzerland},\ \Eprint {https://arxiv.org/abs/2510.02740} {arXiv:2510.02740 [astro-ph.HE]} \BibitemShut {NoStop}%
\bibitem [{\citenamefont {{Bergman}}\ \emph {et~al.}(2025)\citenamefont {{Bergman}} \emph {et~al.}}]{2025arXiv250905530B}%
  \BibitemOpen
  \bibfield  {author} {\bibinfo {author} {\bibfnamefont {D.~R.}\ \bibnamefont {{Bergman}}} \emph {et~al.},\ }\href@noop {} {\bibfield  {journal} {\bibinfo  {journal} {arXiv:2509.05530}\ } (\bibinfo {year} {2025})}\BibitemShut {NoStop}%
\bibitem [{\citenamefont {{Peters}}(1961)}]{1961NCim...22..800P}%
  \BibitemOpen
  \bibfield  {author} {\bibinfo {author} {\bibfnamefont {B.}~\bibnamefont {{Peters}}},\ }\href {https://doi.org/10.1007/BF02783106} {\bibfield  {journal} {\bibinfo  {journal} {Il Nuovo Cimento}\ }\textbf {\bibinfo {volume} {22}},\ \bibinfo {pages} {800} (\bibinfo {year} {1961})}\BibitemShut {NoStop}%
\bibitem [{\citenamefont {{OpenAI}}(2022)}]{openai2022chatgpt}%
  \BibitemOpen
  \bibfield  {author} {\bibinfo {author} {\bibnamefont {{OpenAI}}},\ }\href@noop {} {\bibinfo {title} {Chatgpt: Optimizing language models for dialogue}},\ \bibinfo {howpublished} {\url{https://openai.com/blog/chatgpt}} (\bibinfo {year} {2022}),\ \bibinfo {note} {accessed 14~Aug~2025}\BibitemShut {NoStop}%
\bibitem [{\citenamefont {{Jarrett}}\ \emph {et~al.}(2023)\citenamefont {{Jarrett}} \emph {et~al.}}]{Jarrett2023}%
  \BibitemOpen
  \bibfield  {author} {\bibinfo {author} {\bibfnamefont {T.~H.}\ \bibnamefont {{Jarrett}}} \emph {et~al.},\ }\href {https://doi.org/10.3847/1538-4357/acb68f} {\bibfield  {journal} {\bibinfo  {journal} {\apj}\ }\textbf {\bibinfo {volume} {946}},\ \bibinfo {eid} {95} (\bibinfo {year} {2023})}\BibitemShut {NoStop}%
\bibitem [{\citenamefont {{Stone}}\ and\ \citenamefont {{Metzger}}(2016)}]{2016MNRAS.455..859S}%
  \BibitemOpen
  \bibfield  {author} {\bibinfo {author} {\bibfnamefont {N.~C.}\ \bibnamefont {{Stone}}}\ and\ \bibinfo {author} {\bibfnamefont {B.~D.}\ \bibnamefont {{Metzger}}},\ }\href {https://doi.org/10.1093/mnras/stv2281} {\bibfield  {journal} {\bibinfo  {journal} {\mnras}\ }\textbf {\bibinfo {volume} {455}},\ \bibinfo {pages} {859} (\bibinfo {year} {2016})}\BibitemShut {NoStop}%
\bibitem [{\citenamefont {{Kormendy}}\ and\ \citenamefont {{Ho}}(2013)}]{KormendyHo2013}%
  \BibitemOpen
  \bibfield  {author} {\bibinfo {author} {\bibfnamefont {J.}~\bibnamefont {{Kormendy}}}\ and\ \bibinfo {author} {\bibfnamefont {L.~C.}\ \bibnamefont {{Ho}}},\ }\href {https://doi.org/10.1146/annurev-astro-082708-101811} {\bibfield  {journal} {\bibinfo  {journal} {\araa}\ }\textbf {\bibinfo {volume} {51}},\ \bibinfo {pages} {511} (\bibinfo {year} {2013})}\BibitemShut {NoStop}%
\bibitem [{\citenamefont {{Reines}}\ and\ \citenamefont {{Volonteri}}(2015)}]{ReinesVolonteri2015}%
  \BibitemOpen
  \bibfield  {author} {\bibinfo {author} {\bibfnamefont {A.~E.}\ \bibnamefont {{Reines}}}\ and\ \bibinfo {author} {\bibfnamefont {M.}~\bibnamefont {{Volonteri}}},\ }\href {https://doi.org/10.1088/0004-637X/813/2/82} {\bibfield  {journal} {\bibinfo  {journal} {\apj}\ }\textbf {\bibinfo {volume} {813}},\ \bibinfo {eid} {82} (\bibinfo {year} {2015})}\BibitemShut {NoStop}%
\bibitem [{\citenamefont {{French}}\ \emph {et~al.}(2016)\citenamefont {{French}}, \citenamefont {{Arcavi}},\ and\ \citenamefont {{Zabludoff}}}]{2016ApJ...818L..21F}%
  \BibitemOpen
  \bibfield  {author} {\bibinfo {author} {\bibfnamefont {K.~D.}\ \bibnamefont {{French}}}, \bibinfo {author} {\bibfnamefont {I.}~\bibnamefont {{Arcavi}}},\ and\ \bibinfo {author} {\bibfnamefont {A.}~\bibnamefont {{Zabludoff}}},\ }\href {https://doi.org/10.3847/2041-8205/818/1/L21} {\bibfield  {journal} {\bibinfo  {journal} {\apjl}\ }\textbf {\bibinfo {volume} {818}},\ \bibinfo {eid} {L21} (\bibinfo {year} {2016})}\BibitemShut {NoStop}%
\bibitem [{\citenamefont {{Horiuchi}}\ \emph {et~al.}(2011)\citenamefont {{Horiuchi}}, \citenamefont {{Beacom}}, \citenamefont {{Kochanek}}, \citenamefont {{Prieto}}, \citenamefont {{Stanek}},\ and\ \citenamefont {{Thompson}}}]{Horiuchi2011CCSNR}%
  \BibitemOpen
  \bibfield  {author} {\bibinfo {author} {\bibfnamefont {S.}~\bibnamefont {{Horiuchi}}}, \bibinfo {author} {\bibfnamefont {J.~F.}\ \bibnamefont {{Beacom}}}, \bibinfo {author} {\bibfnamefont {C.~S.}\ \bibnamefont {{Kochanek}}}, \bibinfo {author} {\bibfnamefont {J.~L.}\ \bibnamefont {{Prieto}}}, \bibinfo {author} {\bibfnamefont {K.~Z.}\ \bibnamefont {{Stanek}}},\ and\ \bibinfo {author} {\bibfnamefont {T.~A.}\ \bibnamefont {{Thompson}}},\ }\href {https://doi.org/10.1088/0004-637X/738/2/154} {\bibfield  {journal} {\bibinfo  {journal} {\apj}\ }\textbf {\bibinfo {volume} {738}},\ \bibinfo {eid} {154} (\bibinfo {year} {2011})}\BibitemShut {NoStop}%
\bibitem [{\citenamefont {{Woosley}}\ and\ \citenamefont {{Bloom}}(2006)}]{Woosley2006LGRB}%
  \BibitemOpen
  \bibfield  {author} {\bibinfo {author} {\bibfnamefont {S.~E.}\ \bibnamefont {{Woosley}}}\ and\ \bibinfo {author} {\bibfnamefont {J.~S.}\ \bibnamefont {{Bloom}}},\ }\href {https://doi.org/10.1146/annurev.astro.43.072103.150558} {\bibfield  {journal} {\bibinfo  {journal} {\araa}\ }\textbf {\bibinfo {volume} {44}},\ \bibinfo {pages} {507} (\bibinfo {year} {2006})}\BibitemShut {NoStop}%
\end{thebibliography}%
\end{document}